# La chirurgie urologique assistée par ordinateur et robot

*J. Troccaz[1]*

Laboratoire TIMC

IN3S – Faculté de Médecine

Domaine de la Merci

38706 La Tronche cedex

jocelyne.troccaz@imag.fr ou JTroccaz@chu-grenoble.fr

tel : 04 56 52 00 06 – fax : 04 56 52 00 55

**Mots clés :** robotique, imagerie, informatique

---

[1] Jocelyne Troccaz est Directeur de Recherche au CNRS et responsable de l'équipe « Gestes Médico-Chirurgicaux Assistés par Ordinateur » (cf. http://www-timc.imag.fr/gmcao) du laboratoire TIMC (Unité Mixte de Recherche Université Joseph Fourier/CNRS).

**RESUME**

Ce papier a pour objectif de faire un état de l'art des outils d'informatique et de robotique à disposition de l'urologue. Il concerne l'aide au diagnostic et l'assistance aux gestes et est articulé en deux grandes parties : l'une s'intéresse à l'apport de la robotique et présente plusieurs systèmes à des stades de développement divers (prototypes de laboratoires, systèmes en cours de validation ou systèmes commercialisés). La seconde partie décrit des outils de fusion d'images et des systèmes de navigation en cours de développement ou d'évaluation. Enfin, nous introduisons quelques travaux sur la simulation informatisée des gestes urologiques avant de conclure.

**INTRODUCTION**

Chirurgie laparoscopique, nouveaux modes de destruction des tumeurs (tels que les ultrasons focalisés, la radiofréquence, la cryoablation, etc.), détection de plus en plus précoce des cancers, multiplication des modalités d'imagerie diagnostique et interventionnelle posent de nouveaux challenges à l'urologue en terme de précision et d'efficacité de son geste. Sensibilisée par la médiatisation de certains systèmes robotiques, l'urologie comme d'autres disciplines découvre le potentiel de certains outils de l'informatique et du traitement d'images dans l'aide au diagnostic, à la planification et à la réalisation du geste ou encore à l'enseignement. Dans ce papier, nous faisons un état de l'art aussi complet que possible sur les différentes approches développées (robotisation, fusion de données, navigation chirurgicale pour l'essentiel). Cet état de l'art présente aussi bien des projets de laboratoires en cours de développement ou d'évaluation que des produits industriels. Deux possibilités s'offraient à nous pour cette présentation ; partir des applications cliniques principales et pour chacune lister les différents types d'assistance développés, ou bien partir des outils proposés aux cliniciens et en décrire quelques applications. C'est plutôt cette deuxième présentation que nous avons choisie pour rendre plus claire et plus explicite la généricité de tels outils.

## 1. ROBOTIQUE ET UROLOGIE

Dans ce cadre général d'évolution des gestes médico-chirurgicaux vers moins d'invasité, plus d'efficacité et une traçabilité meilleure, le robot a une place potentielle importante. Historiquement, l'urologie fut un des premiers domaines cliniques où un robot fut expérimenté. En collaboration avec la London Clinic, l'Imperial College of Science, Technology and Medicine de Londres mit au point à partir de 1989 un robot d'assistance à la résection endo-urétrale de prostate [8] appelé PROBOT. Le premier test sur patient eut lieu en Avril 1991. Une faisabilité fut établie à partir de 5 patients puis une étude pré-clinique fut faite sur une série de 40 patients. Différentes versions de ce système furent réalisées car le premier prototype à base de robot industriel (Puma 560 d'Unimation connecté à un cadre limiteur de mouvement) n'était pas propre à un usage clinique. Le système actuel est constitué d'un porteur passif permettant de placer un cadre circulaire supportant 3 degrés de liberté[2] (2 rotations et une translation pour positionner l'outil). Une des limitations de ce système rapportée dans [27] concerne la difficulté à piloter ce robot automatiquement à partir des données ultrasonores dans le but de monitorer la progression de la résection. Nous allons voir dans ce qui suit, que ce robot couplé à une imagerie est typique d'une des deux catégories de robots développés.

En effet, on peut distinguer deux grands types de systèmes robotisés. Dans le premier cas, le robot est couplé à une modalité d'imagerie (scanner, IRM, échographie, fluoroscopie, etc.). Une cible et une trajectoire sont désignées sur l'imagerie et le robot amène un outil (aiguille par exemple) sur cette cible selon cette trajectoire ; le couplage du robot à l'imageur est physique ou se fait par le biais d'un objet de calibration porté par le robot et visible par l'imageur. Nous appelons cette première catégorie

---
[2] Un degré de liberté (ddl) spécifie un paramètre de mouvement : ainsi, le coude a un ddl en rotation, le poignet a 2 ddl principaux en rotation. Un solide libre dans l'espace a 6 ddl (3 translations et 3 rotations).

« *robot guidé par l'image* ». Dans la seconde catégorie de systèmes, le robot perd de son autonomie par le fait qu'il est contrôlé à distance par un opérateur et qu'il répète quasiment à l'identique le geste qui lui est transmis ; il peut cependant être utilisé pour des tâches plus complexes que le robot de première catégorie. Le contrôle à distance d'un robot est appelé télé-opération dans le vocabulaire de la robotique. On parlera donc dans ce qui suit de « *robot télé-opéré* ».

*1.1 Robots couplés à l'imagerie*

1.1.a Biopsies et curiethérapie de la prostate

Beaucoup plus récemment, différentes équipes se sont intéressées au développement de robots facilitant la réalisation de biopsies prostatiques ou le placement de grains radioactifs en curiethérapie. D'un point de vue strictement technique, ces gestes sont très semblables en ce qu'il s'agit dans les deux cas d'insérer une ou plusieurs aiguilles dans la prostate par voie transrectale ou transpérinéale sous contrôle de l'imagerie (le plus souvent de l'échographie endorectale).

En curiethérapie, le système conventionnel intègre mécaniquement la sonde échographique endorectale, un stepper permettant un balayage axial de la glande par les ultrasons et une grille permettant le repérage des aiguilles relativement aux données échographiques et la réalisation du planning dosimétrique. Sous réserve d'une mise en relation mécanique ou logicielle entre le robot et l'imagerie, celui-ci peut se substituer à la grille et dans certains cas permettre des abords selon des trajectoires variées évitant par exemple les conflits avec l'arc pubien. Le planning dosimétrique est dans ce dernier cas plus complexe. On peut citer [9] qui propose un robot simple reproduisant pour l'essentiel les ddl de la grille : deux ddl en translation « axiale » pour le repérage de l'aiguille et un ddl en translation pour sa progression. Une rotation autour de l'axe de l'aiguille est ajoutée afin de réduire les effets de torsion de l'aiguille lors de son insertion dans les tissus. Ce système est à l'état de prototype de recherche. [34] proposent d'utiliser un robot à six ddl permettant de donner n'importe quelle incidence à chacune des aiguilles. La sonde échographique est quant à elle déplacée en rotation autour de son axe pour la reconstruction d'un volume 3D. Des tests ont été effectués sur fantôme. Les problèmes rencontrés par ces systèmes sont communs à la curiethérapie conventionnelle : question de la validité du planning à cause du déplacement des tissus, de l'œdème, flexion des aiguilles. La valeur ajoutée clinique n'est pas très claire surtout mise en rapport avec la complexité de l'introduction du robot dans une procédure chirurgicale.

Afin de faciliter la réalisation des biopsies en lien avec l'imagerie, une approche de robotisation sous IRM est proposée par [31] depuis plusieurs années. L'IRM permet en effet de visualiser clairement la prostate et ses différentes zones. Le robot est réalisé dans des matériaux « transparents » à l'IRM. Seul un micro-aimant associé au guide d'aiguille permet sa visualisation dans l'image IRM. Le patient est installé dans l'IRM et le robot est introduit dans le rectum du patient. Le système a trois ddl : une translation dans le rectum, une rotation autour de son axe principal et la translation de l'aiguille. Après acquisition IRM, la cible est désignée et le robot est commandé pour l'atteindre. [31] présente les principes de ce système ainsi que son évaluation sur des chiens. Les auteurs discutent également la possibilité d'utiliser ce même système pour le placement des grains en curiethérapie sur la base d'un test effectué sur un chien. Comme en témoignent les illustrations du site web [11], le système a été testé sur des patients. Il est évident que le recours à l'IRM pour les biopsies n'est pas envisageable de façon standard ; cependant, pour les patients ayant plusieurs séries de biopsies échoguidées négatives malgré un PSA croissant cette approche peut être de grand intérêt. Plus récemment, [24] ont développé un robot à 9 ddl permettant à la fois de déplacer une sonde échographique endorectale et de positionner une aiguille de biopsie par voie transpérinéale selon un planning préparé à partir des données échographiques – un point d'insertion unique des aiguilles dans le périnée est choisi pour la réalisation de la série des biopsies; des essais cliniques préliminaires rapportent une précision de l'ordre de 2.5mm dans le positionnement de l'extrémité de l'aiguille.

1.1.b Accès rénal percutané

Il s'agit d'atteindre des cibles de dimensions petite à moyenne (typiquement de 2 à 10mm) dans un environnement anatomique contraint, de façon mini-invasive. Ce type de geste se fait traditionnellement sous échographie ou fluoroscopie ; l'échographie donne souvent une mauvaise image de la cible et contraint la trajectoire qui se fait dans le plan de l'image ; cette technique demande une expertise de l'opérateur lui permettant d'apprécier l'information 3D à partir de l'image elle-même bi-dimensionnelle. La fluoroscopie, image projective également bi-dimensionnelle pose par ailleurs le problème de l'irradiation. La morbidité de ce geste est de 5% environ. L'équipe de Johns Hopkins Hospital de Baltimore a développé depuis 1996 un robot nommé PAKY (Percutaneous Access of the KidneY) pour assister ce type de gestes et en améliorer la précision. Le robot à sept ddl passifs (non motorisés) positionne un système actif à trois ddl (motorisés) qui permet l'orientation de l'outil selon deux ddl et l'introduction automatique de l'aiguille selon un troisième ddl en translation. La fluoroscopie est utilisée pour préparer l'alignement de l'aiguille et en contrôler la progression. [7] décrit les expérimentations in vitro et in vivo. Sur une série de 23 patients décrite dans [30], aucune différence significative n'est rapportée entre les procédures robotisée et manuelle en terme de précision, de rapidité, de nombre de tentatives d'accès, de complications. Les auteurs en concluent que la procédure robotisée est faisable et qu'une complète automatisation est envisageable. Ce même robot PAKY est testé pour le placement d'aiguilles de biopsies sous scanner.

*1.2 Robots télé-opérés*

Le recours à une chirurgie laparoscopique en urologie comme dans les autres disciplines cliniques limite les capacités perceptives et motrices du chirurgien. La manipulation de l'endoscope par un assistant mobilise un personnel qualifié et la coordination entre le chirurgien et son aide doit être excellente. Différents outils robotisés ont vu le jour depuis plusieurs années pour pallier à ces difficultés.

1.2.a Robot porte-endoscope

Les premiers travaux concernaient la robotisation des déplacements de l'endoscope. AESOP [26], développé par Computer Motion Inc., a connu un essor important. Ce fut le premier robot approuvé par la FDA aux Etats-Unis dans le contexte de la robotique médicale. Distribué depuis 1994, on comptait en 2000 environ 500 hôpitaux équipés et 100000 procédures réalisées avec son assistance. Ce robot a 4 ddl actifs et 2 ddl passifs et est contrôlé par le chirurgien par commande vocale. Le CEDIT [10] publie une intéressante synthèse concernant l'évaluation clinique de cet assistant sur le plan des complications, du temps d'intervention, du coût et de critères plus subjectifs concernant le confort du chirurgien ; la question de la complexité de l'évaluation apparaît très clairement au travers de ce rapport.

Parallèlement, de nombreux systèmes concurrents ont été étudiés et proposés dans des laboratoires ou des entreprises. L'un d'eux développé par notre laboratoire a pour objectif d'être léger et beaucoup plus facile à intégrer dans un contexte clinique qu'un robot de type « industriel ».

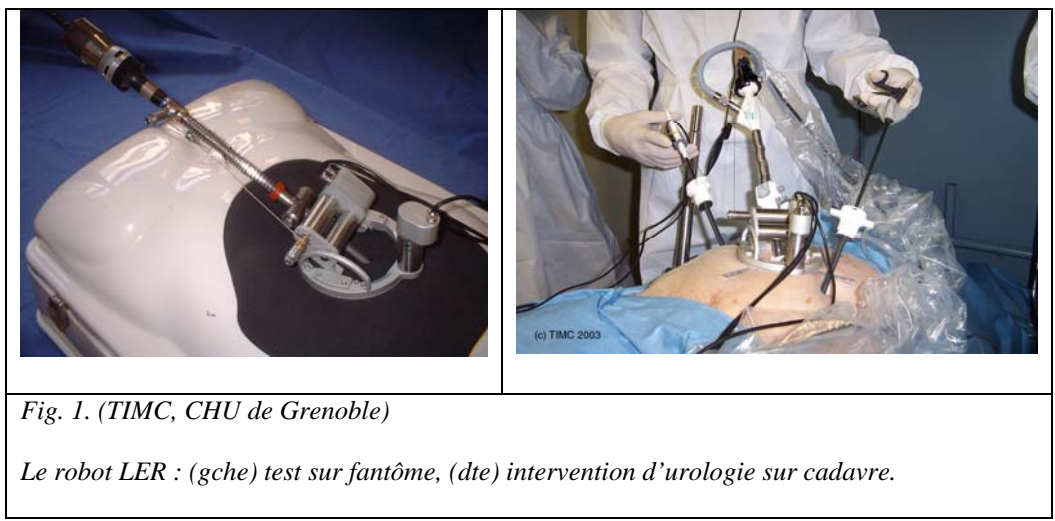

*Fig. 1. (TIMC, CHU de Grenoble)*
*Le robot LER : (gche) test sur fantôme, (dte) intervention d'urologie sur cadavre.*

Le LER [4,16] résultant de ces recherches, est positionné sur le corps du patient (cf. figure 1); il a 3 ddl[3] (les deux rotations de l'outil autour du point d'incision et la translation dans le corps du patient), est complètement stérilisable et pèse 625gr. Il est commandé à la voix par le chirurgien. Son évaluation sur cadavre et sur cochon a permis l'évolution des différents prototypes. La validation clinique en est prévue en 2006 en chirurgie digestive et urologie.

1.2.b Robot de télé-chirurgie

Ces dernières années ont vu l'apparition sur le marché des robots d'assistance au geste de chirurgie endoscopique lui-même et un véritable engouement pour ces systèmes. Le système ZEUS de

---
[3] En effet, contrairement à AESOP positionné sur la table d'opération, le LER est centré sur le point d'incision et 3 degrés de liberté sont suffisants. AESOP quant à lui en nécessite 3 de plus pour le placement relativement à ce point d'incision.

Computer Motion, évolution de l'AESOP, est constitué de l'association de 3 robots dont l'un porte l'endoscope et les deux autres déplacent les instruments sous contrôle du chirurgien assis à un poste maître, situé à quelques mètres de là ; le chirurgien disposant de deux bras maîtres contrôlant les instruments ; le bras porte-endoscope étant contrôlé à la voix comme l'AESOP. Le robot DaVinci, d'Intuitive Surgical Inc., basé sur des principes similaires, dispose quant à lui de ddl intracorporels – en d'autres termes l'extrémité des instruments est elle-même robotisée (cf. figure 2). C'est un avantage indéniable par rapport au système ZEUS.

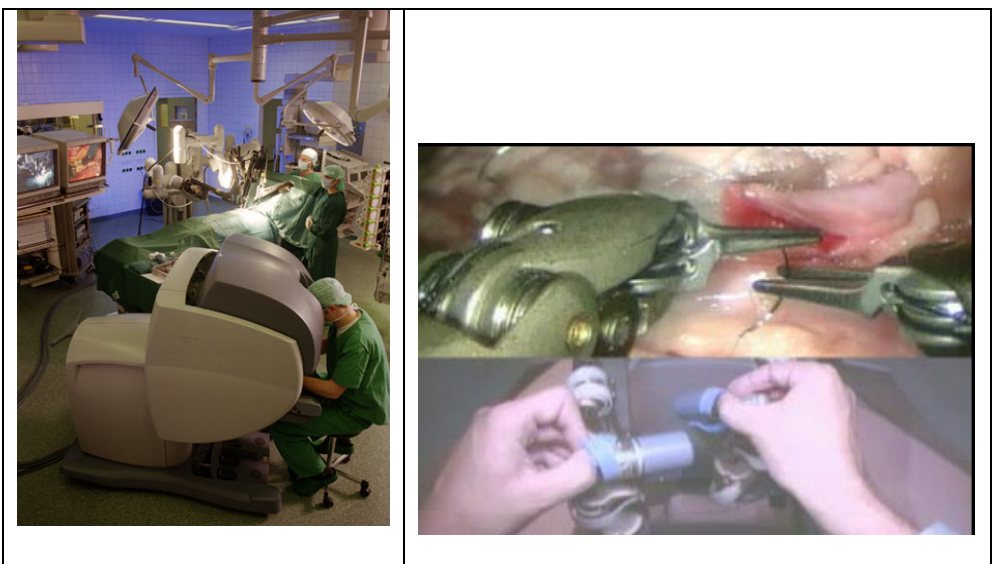

*Fig. 2. (Intuitive Surgical Inc.)*

*Robot DaVinci : (gche) general set-up ; (dte-bas) bras maîtres; (dte-haut) instruments articulés.*

Comme nous le voyons sur la figure 2, le chirurgien est assis au poste maître où il dispose d'une vision stéréoscopique et utilise deux bras maîtres pour déplacer très facilement endoscope et instruments. Ainsi, on combine certains des avantages de la chirurgie laparoscopique et de la chirurgie conventionnelle. Au contraire de ZEUS qui est constitué de 3 robots, DaVinci comporte un pied unique équipé de trois bras et un certain nombre de ddl passifs pour pré-positionner les trois bras relativement aux points d'incision. Le système est en lui-même extrêmement encombrant. Le placement des points d'incision est déterminant pour l'utilisabilité des trois bras et l'accès aux cibles anatomiques. DaVinci a été testé et évalué dans de très nombreuses spécialités dont l'urologie. En urologie, les premières prostatectomies radicales effectuées avec le DaVinci datent de 2000 [1,5]. Depuis, certains centres en ont fait une évaluation extensive sur de grandes séries. Ainsi le Vattikuti Institute de l'hôpital Henry Ford de Détroit aux Etats-Unis publie dans [20] une observation sur plus de 1100 cas. Les prostatectomies radicales laparoscopiques débutèrent en octobre 2000 et l'assistance robotisée avec le DaVinci en mars 2001. Des études comparatives entre chirurgie conventionnelle, laparoscopique et laparoscopique robotisée donnent avantage à ce dernier type d'approche sur de nombreux plans (temps d'hospitalisation, douleurs, pertes sanguines, contrôle du PSA, marges positives, continence, vie sexuelle, etc.).

D'autres applications de ce type de technologie concernent les autres cibles cliniques de l'urologie telles que rein et vessie : on rapporte des adrénalectomies, pyeloplasties, prélèvement de rein sur donneur vivant, cystectomies, etc. effectués avec l'aide d'un tel système de télé-robotique. Nous invitons le lecteur intéressé à se reporter au numéro spécial de Urologic Clinics in North America publié en novembre 2004 (volume 31, numéro 4) et relatif à « Robotic urologic surgery » pour des rapports détaillés sur ces autres applications cliniques.

En 2005, on dénombrait environ 300 systèmes DaVinci installés de par le monde, toutes spécialités utilisatrices confondues, dont environ 50 en Europe et de l'ordre d'une dizaine en France. La question du coût de tels systèmes se pose : ainsi, pour le DaVinci, à l'investissement initial doivent être ajoutés un coût de maintenance annuel très significatif (10% coût initial) et un surcoût sensible par opération. Une évaluation précise coût/bénéfices doit être effectuée pour confirmer l'apport d'un tel système en dehors de tout effet de mode.

De nombreux projets concurrents sont en développement dans les laboratoires visant pour l'essentiel à reproduire les capacités du DaVinci mais en miniaturisant le système. La question du retour d'effort est également à l'étude car le contrôle à distance des bras robotisés par le chirurgien sur la seule base de la vision reste un facteur limitant des dispositifs commerciaux actuels.

1.2.c Télé-mentoring et télé-chirurgie de longue distance

Ces robots permettent à un opérateur distant de déplacer un endoscope ou un instrument. Dans leur usage le plus conventionnel, l'opérateur est dans la même salle que le robot esclave ou dans une salle voisine. Cependant, on en perçoit immédiatement d'autres utilisations possibles en situant l'opérateur à plus grande distance. Ce dernier peut ainsi intervenir comme assistant expert dans une approche de télé-mentoring [17] ou comme réel opérateur dans une approche de télé-chirurgie. L'opération Lindberg [19] rapporte ainsi l'opération depuis New-York, grâce à une variante du système ZEUS et à un système de télécommunication transatlantique dédié, d'une patiente située à Strasbourg pour une cholécystectomie. De telles approches, quoi que prometteuses, posent de nombreux problèmes techniques et légaux.

## 2. FUSION D'IMAGES, NAVIGATION CHIRURGICALE ET UROLOGIE

La robotique a été fort médiatisée dans le domaine médical mais d'autres modes d'assistance existent et peuvent être d'un rapport coût/bénéfice comparable sinon meilleur. Nous présenterons donc ici deux autres types d'assistance : la fusion d'images et la navigation chirurgicale. Leur introduction en clinique peut s'avérer beaucoup plus simple que celle de systèmes robotisés.

*2.1 Fusion d'images*

Comme nous l'avons évoqué précédemment, curiethérapie et biopsie de la prostate utilisent l'échographie transrectale (TRUS). Cette modalité qui permet de visualiser la glande en temps réel de façon non invasive souffre cependant d'une qualité d'images très aléatoire (bruit, artefacts, ombres acoustiques, échogénéïcité du patient) et d'une variabilité intra- et inter-opérateur très importante pour ce qui concerne l'exploitation de ces images. De plus c'est une modalité essentiellement bi-dimensionnelle utilisée dans ce cas pour guider un geste tri-dimensionnel complexe (série de trajectoires à réaliser). Différents travaux ont été développés dans le sens d'un enrichissement de l'échographie par une fusion multi-modale.

[15] présente la fusion de données IRM/TRUS pour le guidage de biopsies transpérinéales. La méthode est basée sur la correspondance de quelques points remarquables repérés en échographie et en IRM. [25] décrit la fusion de données IRM/TRUS pour l'aide à la définition du volume échographique en curiethérapie. Le volume IRM est acquis dans les 3 plans avec une séquence T2 turbo spin echo dans une IRM 1.5T avec antenne endorectale. La méthode repose sur une superposition automatique de la surface de la prostate isolée sur l'IRM et sur l'échographie (cf. figure 3). Les tests effectués avec le système PROCUR (PROstate CURiethérapie) développé sur ce principe comparent le volume échographique obtenu avec ou sans fusion de données. Les résultats obtenus sur 15 patients vont dans le sens d'une sous-estimation du volume échographique sans fusion, pouvant s'avérer significative en terme de l'impact dosimétrique potentiel. L'évaluation de cet impact potentiel est en cours.

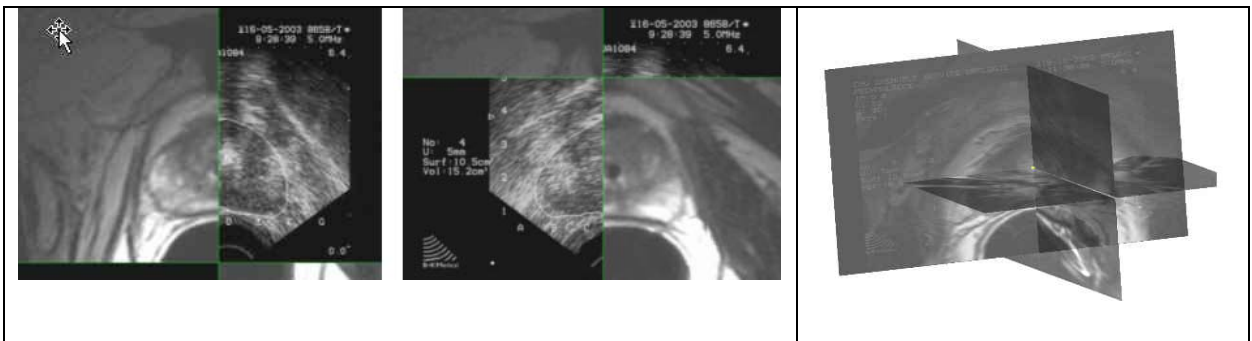

*Fig. 3. (TIMC et CHU de Grenoble)*

*Fusion IRM/TRUS en curiethérapie (PROCUR): (gche, centre) images composites avec lesquelles l'urologue peut interagir – les quadrants haut-gche et bas-droit présentent les données IRM correspondant à l'image TRUS présentée dans les deux autres quadrants ; (dte) représentation 3D avec données IRM dans les 3 plans pour une direction d'aiguille spécifiée – cela permet d'apprécier les rapports de l'aiguille avec son environnement anatomique.*

L'équipe de Harvard Medical School a pour sa part développé des outils de fusion d'images dans le contexte de biopsies transpérinéales effectuées sous IRM interventionnelle [13]. L'idée est d'utiliser ce type d'imagerie pour planifier et réaliser les biopsies. Les images temps réel de l'IRM ouverte sont cependant de qualité moindre; l'approche est donc de fusionner des données d'IRM conventionnelle avec l'IRM interventionnelle. Les patients ont un examen pré-opératoire IRM conventionnel avec des séquences T2 en turbo spin écho dans une machine 1.5T. La procédure de biopsie est quant à elle réalisée dans une IRM ouverte 0.5T. Un volume est également acquis dans les conditions per-opératoires avant biopsie. En pré-opératoire comme en per-opératoire, une antenne externe pelvienne est utilisée. Une grille de guidage des curiethérapies sert de système de guidage des aiguilles de biopsie; les coordonnées de la grille sont localisées dans l'IRM grâce à une calibration. Après fusion de données automatique entre les données pré-opératoires et per-opératoires, les zones d'intérêt (zone périphérique, éventuelle zones suspectes) sont connues dans le référentiel per-opératoire. Des images temps réel (fast gradient recalled-echo) en T2 sont acquises au rythme d'environ une image toutes les 8s pour contrôler la progression de l'aiguille et ces images sont associées aux coupes correspondantes recalculées dans le volume per-opératoire. Ces deux types d'images (T2 turbo spin écho et T2 fast gradient) sont très complémentaires pour le guidage. [13] présente les expérimentations avec deux patients. La même technique peut être utilisée pour la curiethérapie. Elle parait très intéressante mais est cependant limitée au peu de centres équipés d'IRM ouvertes. Evidemment, comme la technique décrite en 1.2, le recours à l'IRM pour les biopsies ne se justifie pas en routine standard et doit être réservée aux patients pour lesquels la procédure conventionnelle n'est pas applicable ou n'est pas satisfaisante.

[3] propose d'utiliser une fusion de données IRM/coupes histologiques pour une meilleure compréhension de l'imagerie IRM du cancer de la prostate. Le logiciel PROPATH réalisant la fusion utilise des données acquises sur des patients traités chirurgicalement pour des cancers de la prostate ; il permettra une étude rétrospective des données.

*2.2 Navigation chirurgicale*

L'idée de navigation chirurgicale a été introduite en neurochirurgie dans les années 80. Il s'agit de suivre la position des instruments par rapport à l'anatomie du patient et à un éventuel planning (cible et trajectoire) et de présenter cette information relativement à une imagerie pré- ou per-opératoire sans que l'instrument n'ait à être directement visible dans les images. Classiquement une imagerie pré-opératoire (scanner ou IRM) est réalisée et sert à planifier l'intervention. En per-opératoire, une imagerie souvent plus « légère » (radiographie, échographie, etc.) est fusionnée avec l'imagerie pré-opératoire permettant le transfert du planning dans les conditions de l'intervention. Les instruments chirurgicaux sont le plus souvent suivis en temps réel dans l'espace grâce à un localisateur. Leur position relativement au planning est visualisée sur un écran permettant au chirurgien de réaliser son geste avec une très grande précision. Ce concept de navigation est à rapprocher de l'idée du GPS permettant la navigation sur la base d'une cartographie (navale ou terrestre) préenregistrée.

2.2.a Ponction rénale

Nous avons développé une telle assistance dans le cas des ponctions rénales percutanées (cf. figure 4) dont la difficulté a été introduite en 1.1.b.

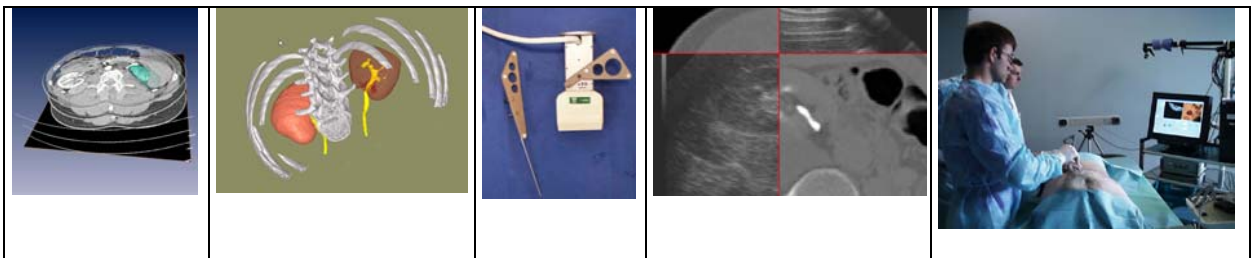

*Fig. 4 (TIMC, Hôpital de la Pitié-Salpétrière, PRAXIM)*

*Navigation de la ponction rénale percutanée : volume scanner et rein segmenté – interface de planning pré-opératoire – sonde échographique et aiguille de ponction localisées dans l'espace – fusion de données scanner/échographie – guidage de la ponction.*

L'approche [18,23] consiste à planifier le geste sur la base de l'examen scanner ; deux volumes avec injection sont acquis aux temps précoce et tardif et sont fusionnés automatiquement par informatique. Le planning est réalisé sur ces données : cible et trajectoire sont choisies. Lors de l'intervention, une échographie localisée dans l'espace est utilisée pour acquérir le rein en conditions opératoires. Chaque coupe échographique étant repérée dans l'espace grâce à un système de localisation, on obtient des données de nature tridimensionnelle. Après fusion automatique des données échographiques avec les données scanner, le planning est transféré dans le référentiel opératoire. L'aiguille de ponction est localisée en temps réel et sa trajectoire est repérée par rapport au planning sur une interface visuelle qu'utilise le clinicien pour se guider vers sa cible. Des tests sur fantôme, sur pièce isolée et sur cadavre sont tout à fait prometteurs. Une approche similaire peut également être appliquée à d'autres gestes sur le rein ou d'autres cibles anatomiques. Ainsi, cette même idée de guidage a été mise en œuvre pour la neuro-modulation de la racine sacrée S3 en cas d'incontinence urinaire. L'approche conventionnelle réalisée sous fluoroscopie reste délicate. Nous avons donc entrepris d'utiliser l'approche de navigation décrite ci-dessus. Les tests sur cadavre avec scanner post-opératoire de contrôle démontrent son potentiel [22].

2.2.b Biopsies de la prostate

La réalisation précise des schémas de biopsies (en sextant par exemple) se heurte à la difficulté à se localiser précisément à partir des images échographiques endorectales bidimensionnelles. L'approche de navigation peut permettre une réalisation plus précise du schéma de biopsies. Ainsi, nous avons équipé une sonde échographique endorectale d'un système de localisation permettant en temps réel de connaître précisément la position de la sonde et du guide de ponction dans un référentiel fixe et relativement à la prostate et de visualiser les biopsies réalisées (cf. figure 5). Il s'agit pour l'instant d'un simple enregistrement des trajectoires de biopsies réalisées [6]. Ce travail en est encore à un stade encore préliminaire. En effet les déplacements et déformations de la prostate sous l'effet des mouvements de la sonde échographique limitent la précision de l'enregistrement des données. Une approche de fusion de données entre une imagerie échographique 3D et une imagerie échographique temps réel 4D est en cours de développement. Le système PRONAV (PROstate NAVigation) résultant de cette approche permettra dans sa version ultérieure de guider le clinicien afin de réaliser précisément le schéma sélectionné ou d'atteindre une cible visualisée sur l'IRM par exemple.

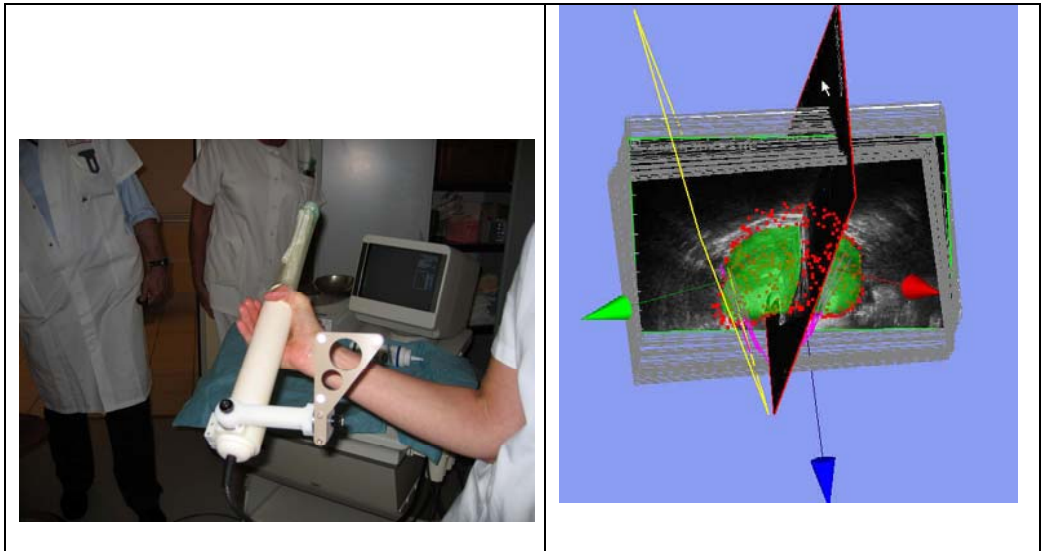

*Fig. 5 (TIMC, CHU de Grenoble)*

*Navigation des biopsies (PRONAV) : Sonde échographique équipée de son système de localisation – interface de visualisation des trajectoires de biopsies réalisées relativement à la prostate (en vert : la prostate – en jaune : un plan échographique dans lequel une biopsie a été réalisée – en rose : les trajectoires de biopsie enregistrées).*

En parallèle de cela, la question du choix du schéma de biopsies reste un sujet très discuté [33]. Afin de faire reposer ce choix sur des données statistiques de localisation des tumeurs dans la prostate, différents travaux visent à élaborer des atlas statistiques à partir de pièces réséquées sur des patients traités chirurgicalement pour des cancers de la prostate. Différents auteurs présentent une méthodologie de préparation et de traitement des coupes histologiques [3] et de construction de l'atlas [29]. Ces derniers proposent également pour chaque nouveau patient, un algorithme optimisant, à partir de l'atlas et des données du patient, le schéma des biopsies pour maximiser le pouvoir diagnostic des biopsies.

## 3. SIMULATEURS CHIRURGICAUX INFORMATISES

La dernière décennie a vu le développement de nombreux outils dédiés à l'enseignement de la chirurgie laparoscopique ; il s'agissait notamment de permettre à un opérateur de s'entraîner à certaines procédures sur un patient virtuel informatisé, en donnant à cet opérateur un retour visuel et tactile réaliste au travers de dispositifs à retour d'effort. L'intérêt de telles approches réside dans la possibilité d'expérimenter sans danger de nouvelles techniques, de permettre de pratiquer des cas rares, de donner des outils quantitatifs pour l'évaluation de l'opérateur. Dans le contexte spécifique de l'urologie, différents simulateurs informatisés ont été proposés entre autres pour le toucher rectal, les résections trans-urétrales de prostate, l'urétéroscopie flexible, les procédures laparoscopiques ; on trouvera dans [28] une bonne introduction à ces travaux. De nombreux travaux actuels concernent le développement de simulateurs pour les biopsies ou les curiethérapies de la prostate (cf. [12,14]). Quelques produits commerciaux sont également disponibles comme c'est le cas pour URO Mentor (cf. [21]).

Un certain nombre de procédures, pour être simulées de façon réaliste, passent par la modélisation fine des interactions des outils (bistouris, aiguilles, endoscopes) avec les tissus vivants; elles nécessitent de résoudre des questions complexes telles que par exemple la modélisation de l'œdème opératoire lors de la curiethérapie ou le saignement lors de procédures endo-urologiques [32].

Certains simulateurs peuvent enfin aider au planning des interventions ; ainsi l'équipe de Créteil a appliqué des outils développés par l'INRIA de Sophia Antipolis pour l'optimisation du placement des ports lors d'une lomboscopie assistée par robot DaVinci (cf. [2]).

**CONCLUSION**

Comme nous l'avons vu dans ces différents chapitres, de nombreuses voies s'offrent aux mondes scientifique et industriel pour assister l'urologue dans les prises en charge diagnostiques ou thérapeutiques de ses patients voire dans l'apprentissage de ces pratiques. Cette nouvelle approche de l'urologie est pourtant relativement récente. En effet, la chirurgie assistée par ordinateur et robot a connu ses premiers développements au milieu des années 80 sur des structures anatomiques rigides (ou considérées comme telles) : neuro-chirurgie, ORL, orthopédie notamment. Pour l'urologie comme pour d'autres spécialités, une difficulté supplémentaire est inhérente aux tissus mous dont la mobilité et la déformabilité pendant l'acquisition d'images ou pendant le geste doivent être prises en compte. La télé-chirurgie robotisée telle que présentée en 1.2.b ne pose pas les mêmes problèmes puisque le chirurgien « ferme la boucle » : il perçoit ces changements en temps réel et adapte son geste en conséquence ; c'est pourquoi des produits ont vu le jour relativement rapidement dans ce domaine. Par contre, le plus souvent, assister informatiquement des gestes sur des organes tels que la prostate ou le rein ou les simuler nécessite d'être capable de modéliser et/ou de suivre ces modifications en temps réel ; c'est un problème difficile impliquant des recherches très pluri-disciplinaires (travaux de modélisation, développement de capteurs, traitement de données temps réel, robotique couplée à l'imagerie, etc.). De nombreux laboratoires ont compris le potentiel scientifique et clinique de ces domaines de recherche et c'est de très bon augure quant à l'émergence de nouveaux outils d'assistance informatisée à l'urologie. L'impact en terme de santé publique peut être conséquent.

**BIBLIOGRAPHIE**